\documentclass[aip,reprint]{revtex4-1}

\usepackage{graphicx}
\usepackage{mathtools}
\usepackage{units}

\begin{document}
\title{Ultrafast magnetization switching by spin-orbit torques}
	
\author{Kevin Garello}
\affiliation{Department of Materials, ETH Z\"{u}rich, H\"{o}nggerbergring 64, CH-8093 Z\"{u}rich, Switzerland}
\email{kevin.garello@mat.ethz.ch}

\author{Can Onur Avci}
\affiliation{Department of Materials, ETH Z\"{u}rich, H\"{o}nggerbergring 64, CH-8093 Z\"{u}rich, Switzerland}

\author{Ioan Mihai Miron}
\affiliation{SPINTEC, UMR-8191, CEA/CNRS/UJF/GINP, INAC, F-38054 Grenoble, France}

\author{Manuel Baumgartner}
\affiliation{Department of Materials, ETH Z\"{u}rich, H\"{o}nggerbergring 64, CH-8093 Z\"{u}rich, Switzerland}

\author{Abhijit Ghosh}
\affiliation{Department of Materials, ETH Z\"{u}rich, H\"{o}nggerbergring 64, CH-8093 Z\"{u}rich, Switzerland}

\author{St\'{e}phane Auffret}
\affiliation{SPINTEC, UMR-8191, CEA/CNRS/UJF/GINP, INAC, F-38054 Grenoble, France}

\author{Olivier Boulle}
\affiliation{SPINTEC, UMR-8191, CEA/CNRS/UJF/GINP, INAC, F-38054 Grenoble, France}

\author{Gilles Gaudin}
\affiliation{SPINTEC, UMR-8191, CEA/CNRS/UJF/GINP, INAC, F-38054 Grenoble, France}

\author{Pietro Gambardella}
\affiliation{Department of Materials, ETH Z\"{u}rich, H\"{o}nggerbergring 64, CH-8093 Z\"{u}rich, Switzerland}

\date{\today}	
\begin{abstract}	
Spin-orbit torques induced by spin Hall and interfacial effects in heavy metal/ferromagnetic bilayers allow for a switching geometry based on in-plane current injection. Using this geometry, we demonstrate deterministic magnetization reversal by current pulses ranging from 180~ps to ms in Pt/Co/AlO$_x$ dots with lateral dimensions of 90~nm. We characterize the switching probability and critical current $I_c$ as function of pulse length, amplitude, and external field. Our data evidence two distinct regimes: a short-time intrinsic regime, where $I_c$ scales linearly with the inverse of the pulse length, and a long-time thermally assisted regime where $I_c$ varies weakly. Both regimes are consistent with magnetization reversal proceeding by nucleation and fast propagation of domains. We find that $I_c$ is a factor 3-4 smaller compared to a single domain model and that the incubation time is negligibly small, which is a hallmark feature of spin-orbit torques.
\end{abstract}
\maketitle
%
%%%%%%%%%%%%%%%%%%%%%%%%%INTRODUCTION%%%%%%%%%%%%%%%%%%%%%%%%%%%%%%%%%%%%
Magnetization switching is a topic of fundamental interest as well as of practical relevance for the development of fast, non-volatile data storage devices. In recent years, current-induced switching of nanosized magnets has emerged as one of the most promising technologies for the realization of a scalable magnetic random access memory (MRAM).~\cite{IRTS} In the so-called spin transfer torque (STT)-MRAM,~\cite{dieny2010spin} a spin-polarized current flowing through a pinned magnetic layer induces a torque on the storage layer that counteracts the magnetic damping.~\cite{slonczewski96jmmm} STT switching can be made faster by increasing the injected current or choosing materials with low damping. However, when the magnetization of the reference and free layer are at rest, parallel or anti-parallel, the STT is zero.
The resulting non-negligible incubation delay, governed by thermally activated oscillations, limits ultrafast switching and induces a broad switching time distribution.~\cite{devolderPRB2007} Several solutions have been explored to reduce the incubation delay, such as biasing STT devices with a hard axis field~\cite{devolderPRB2007} or adding an out-of-plane polarizer to an in-plane free layer.~\cite{kentAPL2004} This has led to switching times as low as 50~ps in metallic spin valves~\cite{papusoiAPL2009,lee11apl} and 500~ps in magnetic tunnel junctions (MTJ).~\cite{liu2010ultrafast} Despite such progress, the development of STT-MRAM for ultrafast applications such as cache memories remains problematic. Fast switching requires passing a large current through the thin oxide barrier of a MTJ, which leads to reliability issues and accelerated aging of the barrier.

\begin{figure}	[b]
	\centering	
	\includegraphics[width=8 cm]{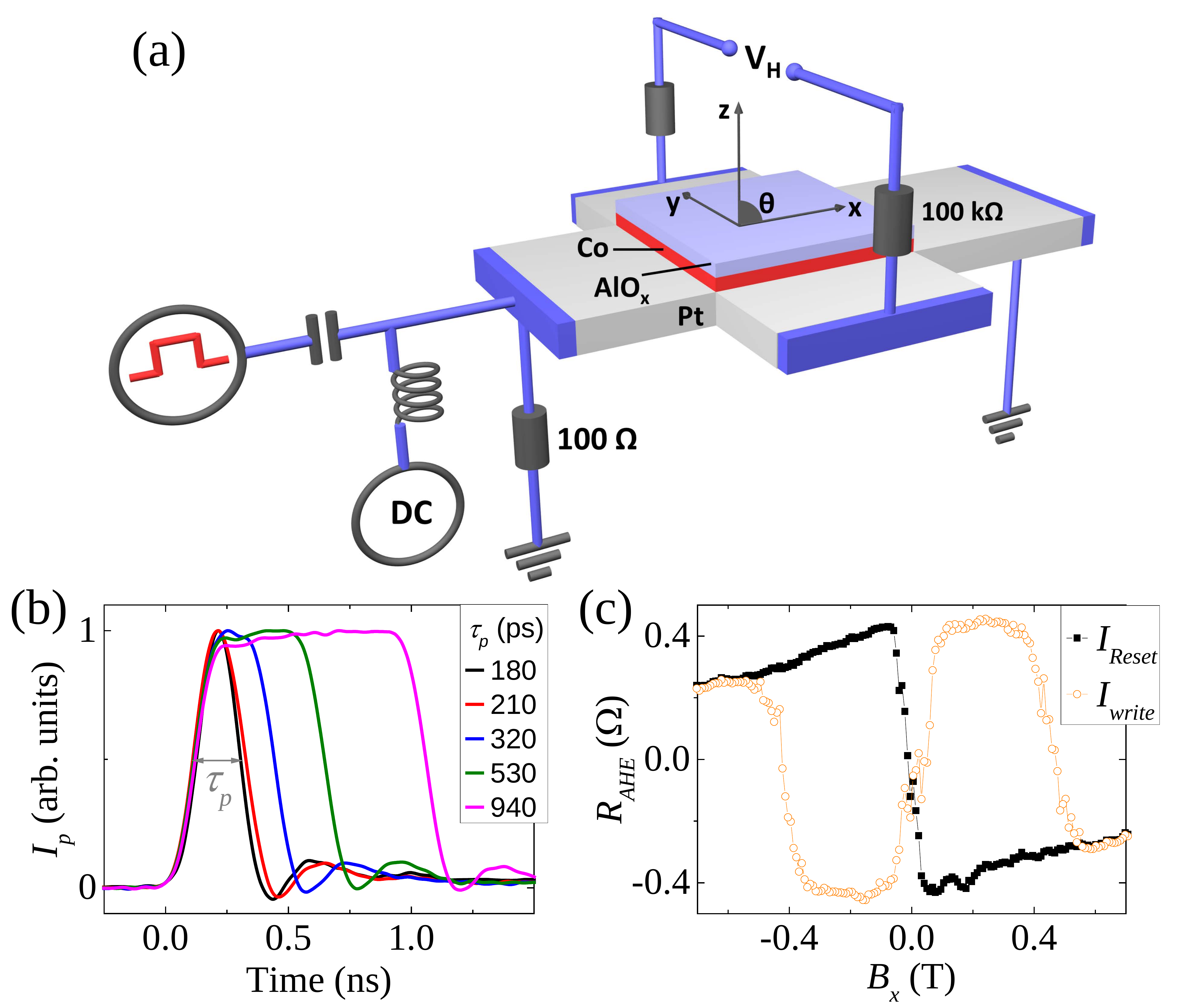}\\	
	\caption{(a) Schematic of the experimental setup. (b) Current pulses of different duration detected in transmission. (c) Magnetization switching of sample s1 induced by positive and negative current pulses with current density $I_p = 1.65$~mA and $\tau_{p}=210$~ps. Note that $B_{x}$ is swept only once from +0.65 to -0.65~T.}\label{fig1}	
\end{figure}
Spin-orbit torque (SOT)-induced switching, generated by the flow of an electrical current in the plane of a ferromagnetic/heavy metal (FM/HM) bilayer, offers an interesting alternative to STT.~\cite{MironN2011} Theoretical~\cite{HaneyPRB2013,wang12prl} and experimental~\cite{MironN2011,MironNM2010,LiuS2012,pai2012APL,GarelloNN2013,KimNM2013,FanNC2014,ferguson14apl} studies have evidenced significant antidamping and field-like SOT components in such systems, which originate from either the bulk spin Hall effect in the HM layer or interfacial Rashba-type spin-orbit coupling, or a combination of these effects. Independently of their origin, SOT have proven very effective to switch the magnetization of perpendicular~\cite{MironN2011,AvciAPL2012,AvciPRB2014} and in-plane magnetized layers,~\cite{LiuS2012,pai2012APL} as well as to control the motion of domain walls in FM/HM heterostructures.~\cite{MironNM2011,emoriNM2013,ryuNNT2013} In SOT devices, as the antidamping torque is always perpendicular to the magnetization, the incubation delay of the switching process is expected to be minimum. Moreover, SOT allow for the separation of the read and write current paths in an MTJ, avoiding electrical stress of the tunnel barrier during writing. Based on these considerations, novel SOT-MRAM architectures have been proposed~\cite{GaudinPatent2012} and the switching of in-plane~\cite{pai2012APL,LiuS2012,yamanouchiAPL2013} and out-of-plane MTJ have been recently demonstrated.~\cite{cubukcuAPL2014} There is however no systematic study of SOT switching on a sub-ns timescale. In this letter, we investigate the probability of SOT-induced magnetization reversal of perpendicularly magnetized Pt/Co/AlO$_x$ dots as a function of current pulse width, amplitude, and external magnetic field on timescales ranging from ms to 180~ps.

%
%%%%%%%%%%%%%%%%%%%%%%%%%%%%%%%%%%%EXPERIMENT%%%%%%%%%%%%%%%%%%%%%%%%%%%%%%%%%%%%%%%%%%%%%%%%%%%%%%%%
\begin{figure}
	\centering	
	\includegraphics[width=8.5 cm]{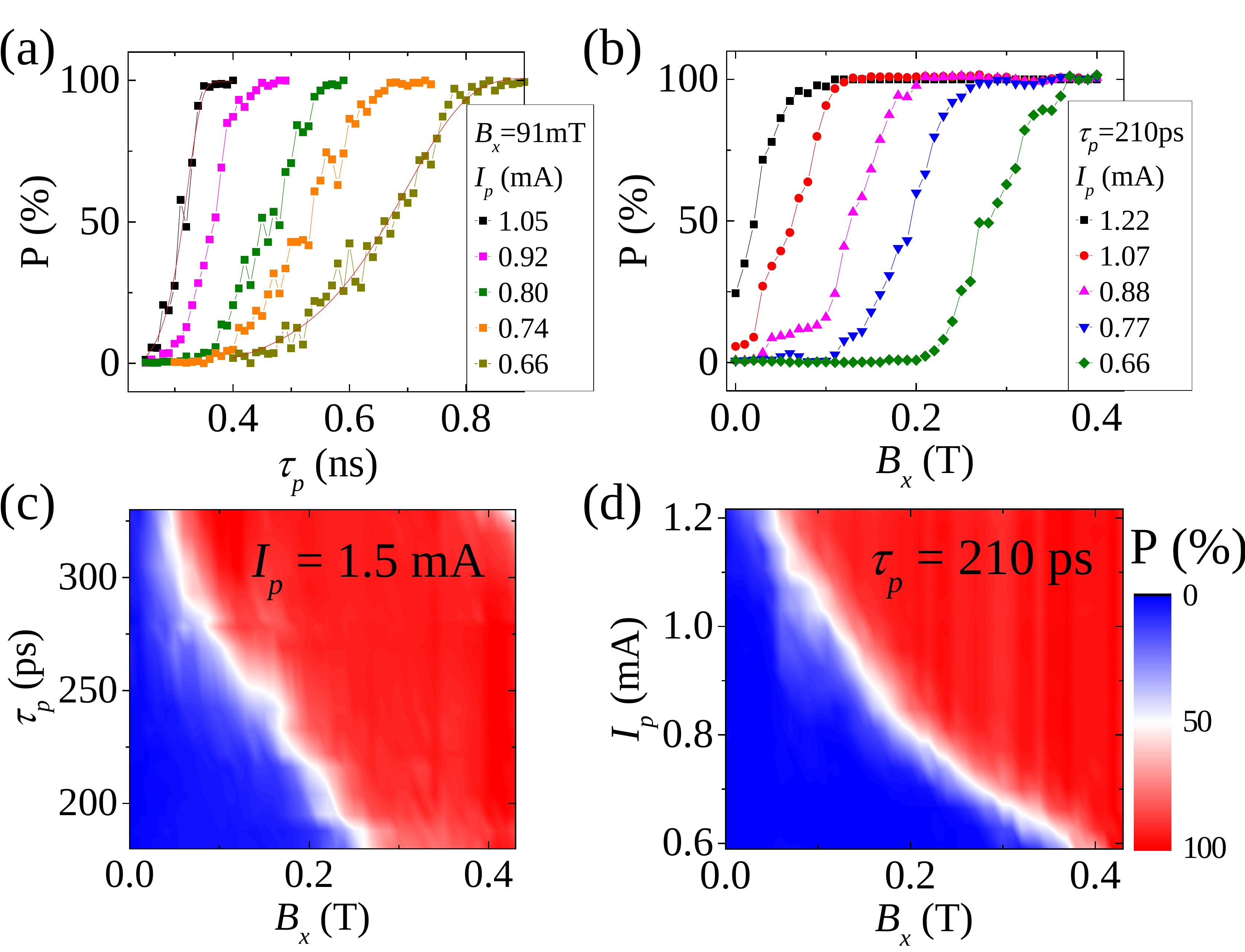}\\
	\caption{Switching probability of s1 as a function of (a) $\tau_p$ ($B_{x}=91$~mT) and (b) $B_{x}$ ($\tau_p=210$~ps) at different current amplitudes. Two-dimensional diagrams of the switching probability showing successful (red) and unsuccessful (blue) events measured as a function of (c) $\tau_p$ and $B_{x}$ for fixed $I_p=1.5$ mA and (d) $I_p$ and $B_{x}$ for fixed $\tau_p=210$~ps.
	}\label{fig2}	
\end{figure}
Pt(3nm)/Co(0.6nm)/AlO$_x$ layers with perpendicular anisotropy were deposited by magnetron sputtering and patterned into square dots on top of Pt Hall bars, as described in Ref.~\onlinecite{MironN2011}. We present results for three different samples of lateral size s1 = 90~nm, s2 = 95~nm and s3 = 102~nm, as measured by scanning electron microscopy. These samples have a saturation magnetization $M_s \approx 8.7 \times 10^5$~A/m (measured before patterning) and an effective anisotropy field $B_k=2K/M_s - \mu_0M_s \approx 1$~T. Figure~\ref{fig1} (a) shows a schematic of the measurement setup. The current pulses are applied along $x$. In order to ensure the transmission of fast pulses without significant reflection due to the large resistance of the Pt contacts ($\sim 2$~k$\Omega$), a 100~$\Omega$ resistor is connected in parallel with the sample. A 100~k$\Omega$ series resistor prevents spreading of the current pulses into the Hall voltage probes. As the antidamping SOT destabilizes both directions of the magnetization, an in-plane bias magnetic field ($B_{x}$) is required to determine the switching polarity.~\cite{MironN2011,AvciAPL2012} Here $B_{x}$ is applied with a tilt of $0.5^\circ$ towards $z$ in order to favor a homogeneous magnetization when no current pulses are applied. The perpendicular component of the magnetization is measured via the anomalous Hall resistance ($R_{AHE}=0.9$~$\Omega$ at saturation) using a low DC current of 20~$\mu$A. A bias tee separates the current pulses and the DC current. All measurements are performed at room temperature. To study the switching probability distribution, we proceed as follows: first a positive 0.7~mA "reset" pulse of 20~ns duration is used to initialize the magnetization direction. Second, a negative "write" pulse of length $\tau_{p}$ and amplitude $I_{p}$ is applied. $R_{AHE}$ is measured a few milliseconds after each pulse. The switching probability is defined as $P=[R_{AHE}(write) - R_{AHE}(reset)] / R_{AHE}$ averaged over 100 trials. Switching diagrams are constructed by varying two out of the three free parameters $\tau_{p}$, $I_{p}$, and $B_{x}$ while the other one is kept constant.

%
%%%%%%%%%%%%%%%%%%%%%%%%%%%%%%%%%%%RESULTS%%%%%%%%%%%%%%%%%%%%%%%%%%%%%%%%%%%%%%%%%%%%%%%%%%%%%%%%
Figure~\ref{fig1} (c) shows the magnetization of sample s1 after applying write pulses with $\tau_{p}$ = 210~ps and $I_{p}$ = 1.65~mA (open orange circles) as a function of $B_{x}$. The magnetization after the reset operation is shown as solid black squares. $B_{x}$ is swept in steps from -650 mT to 650 mT. At each field step, $R_{AHE}$ is measured  after each pulse and averaged over 100 pulses. In the hysteretic range delimited by the coercivity of the Co layer, the orange and black curves indicate that for $B_{x}>0$ a current $I_{p} > 0$ switches the magnetization downwards and $I_{p} < 0$ switches it upwards, whereas for $B_{x}<0$ the effect of the current polarity is reversed. This behavior is typical of SOT and similar to that reported for single pulses ranging from tens of ns to $\mu s$ in devices with size varying from 200 to 1000~nm.~\cite{MironN2011,AvciAPL2012,AvciPRB2014,cubukcuAPL2014}

Since switching occurs on such short timescales and considering the analogy between orthogonal-STT devices and SOT (polarization of the spin current  perpendicular to the magnetization), effects related to the magnetization precession are expected to be important when varying $\tau_{p}$ and $I_{p}$.~\cite{papusoiAPL2009,MarinJAP2012} Moreover, macrospin simulations show that the field-like SOT component (equivalent to an effective field along $y$) promotes oscillations of the magnetization with periods up to ns, thus inducing precessional switching even for high damping constants such as $\alpha=0.5$. We therefore measured the switching probability as a function of $\tau_{p}$ and $I_{p}$, as well as of $B_{x}$, which plays an important role in the SOT-induced dynamics. Figures~\ref{fig2} (a) and (b) show representative measurements of $P$ as a function of $\tau_{p}$ and $B_{x}$, respectively, for different values of $I_p$. By repeating such measurements over a grid of ($B_{x}$, 	$\tau_{p}$) and ($B_{x}$, $I_{p}$) pairs, we construct the switching diagrams reported in Figures~\ref{fig2} (c) and (d). The red (blue) color represents successful (unsuccessful) switching. In both diagrams, the range of successful switching events grows monotonically as either $I_{p}$, $\tau_{p}$ or $B_{x}$ increase. We observe that the white boundary region representing intermediate $P$ values is relatively narrow and that $P$ does not oscillate beyond this boundary, as would be expected for precessional switching.~\cite{papusoiAPL2009,MarinJAP2012} This implies that SOT-induced magnetization reversal in our samples is deterministic and bipolar with respect to either field or current down to $\tau_{p} = 180$~ps.

\begin{figure}
	\centering	
	\includegraphics[width=7.5 cm]{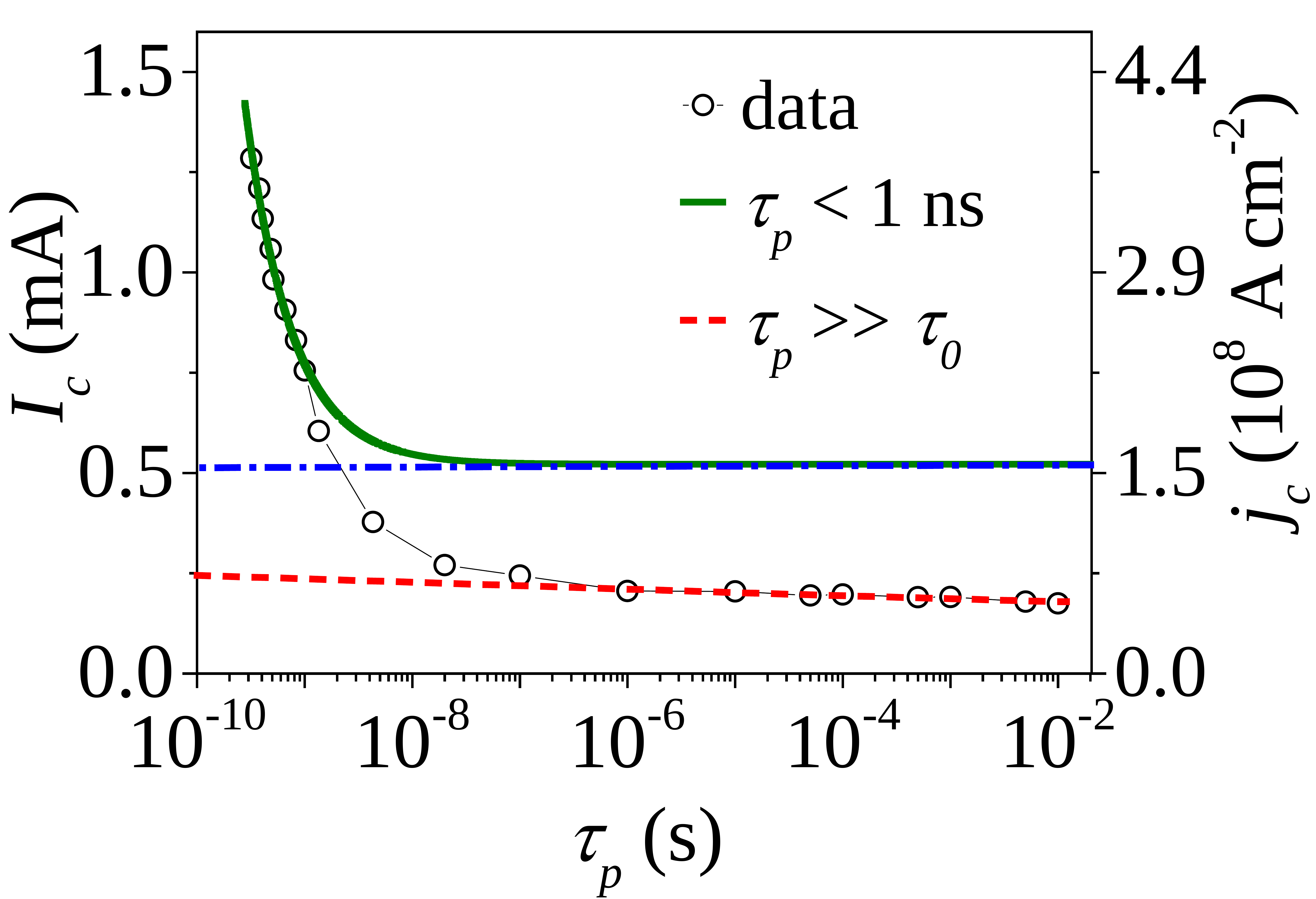}\\	
	\caption{Critical switching current of sample s2 as a function of pulse duration measured with $B_{x}$ = 91 mT. The green solid line is a fit to the data in the short-time regime ($\tau_{p} < 1$~ns) according to Eq.~\eqref{eq1}. The red dashed line is a fit to the data in the thermally activated regime ($\tau_{p}\geq 1$~$\mu$s) according to Eq.~\eqref{eq2}. The blue dash-dotted line represents the intrinsic critical current $I_{c0}$.}\label{fig3}
\end{figure}
$I_p$ and $\tau_{p}$ determine the energy dissipation during the switching process and the speed at which this can be achieved for a given bias field. Figure~\ref{fig3} shows the critical switching current $I_c$, defined at $P=90$~\%, as a function of $\tau_{p}$ measured over eight orders of magnitude in pulse duration for $B_{x} = 91$~mT. We find that there are two very different regimes: at short-time scales ($\tau_{p}< 1$~ns) $I_c$ increase strongly when reducing $\tau_{p}$, whereas on longer time scales ($\tau_{p}\geq 1$~$\mu$s), $I_c$ has a weak dependence on $\tau_{p}$. This behavior is qualitatively similar to that observed in STT devices~\cite{kochPRL2004,bedauAPL2010,liuJMMM2014} and associated with an intrinsic regime where the switching speed depends on the efficiency of angular momentum transfer from the current to the magnetic layer and a thermally assisted regime in which stochastic fluctuations help the magnetization to overcome the reversal energy barrier.

\begin{figure}	
	\centering	
	\includegraphics[width=8 cm]{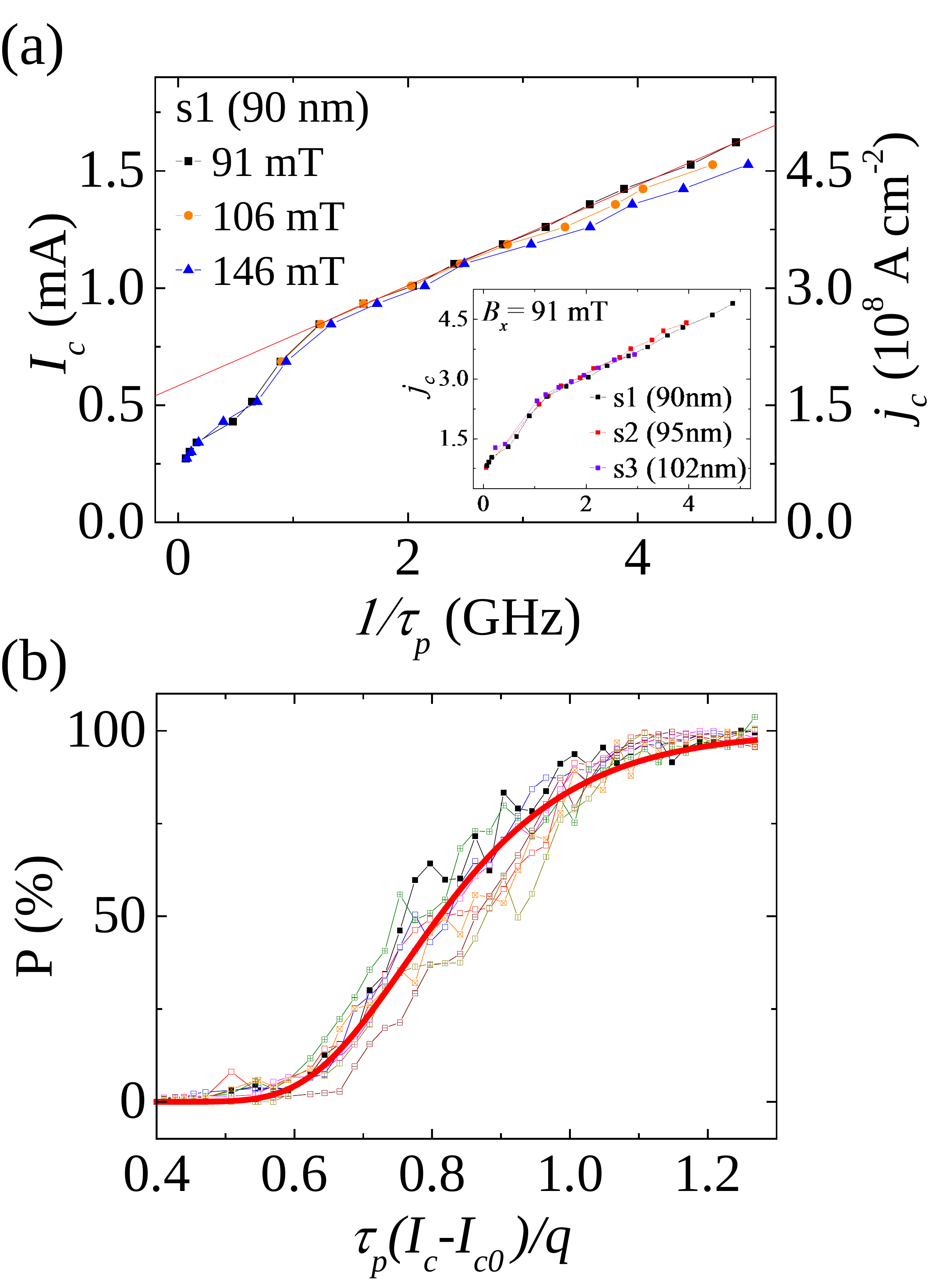}\\	
	\caption{(a) Critical switching current of sample s1 as a function of $1/\tau_{p}$ for different values of $B_x$. The thin red line shows a linear fit to the short-time data ($1/\tau_{p} > 1$~ns) measured at $B_{x}=91$ mT using Eq.~\eqref{eq1}. Inset: similar behavior observed on different samples. The s1, s2, and s3 curves superimpose once normalized to the critical current density $j_c$. (b) Switching probability in the short-time regime as a function of the adimensional parameter $\tau_p(I_c-I_{c0})/q$. The red line represent an average fit of all the curves using a sigmoidal function.}\label{fig4}	
\end{figure}
We focus first on the short-time regime. In this limit, $I_c$ is inversely proportional to $\tau_{p}$, as shown in Figure~\ref{fig4}(a). Similar behavior is observed for samples s1-s3 [inset of Fig.~\ref{fig4}(a)] as well as in larger dots (not shown), indicating that the $\tau_{p}^{-1}$ dependence is specific to the switching process rather than to a particular sample. In analogy with STT,~\cite{bedauAPL2010,liuJMMM2014} we model $I_c$ as
\begin{equation}
\label{eq1}
I_c = I_{c0} + \frac{q}{\tau_p} \, ,
\end{equation}
where $I_{c0}$ is the intrinsic critical switching current and $q$ is an effective charge that describes the efficiency of angular momentum transfer from the current to the spin system. For the fit shown in Fig.~\ref{fig4}(a) we obtain $I_{c0} = 0.58$~mA ($j_{c0} = 1.76\times 10^{8}$~Acm$^{-2}$) and $q = 2.1 \times 10^{-13}$~C. This linear relationship holds for different $B_{x}$ [Fig.~\ref{fig4}(a)]. When increasing $B_{x}$ from 91 to 146~mT $q$ decreases by about 13~\%, whereas $I_{c0}$ increases from 0.58 to 0.61~mA. Further proof that the linear dependence of $I_{c}$ on $\tau_{p}^{-1}$ is general to the switching distribution and not dependent on the definition of the critical current is reported in Fig.~\ref{fig4}(b), showing that all the switching probability curves measured for $\tau_{p}<1$~ns, plotted as a function of the scaled angular momentum $(I_c-I_{c0})\tau_p /q$, fall onto the same curve.

The intrinsic threshold current $I_{c0}$ for SOT switching of a perpendicular layer has been calculated analytically using a monodomain approximation by equating the antidamping SOT to the magnetic anisotropy energy barrier modified by the external field,~\cite{LeeAPL2013} namely, by setting $T^{\parallel}(I_{c0})=(B_k/2-B_x/\sqrt{2})$, where $T^{\parallel}$ is the amplitude of the antidamping torque per unit of magnetic moment. This torque is often expressed in terms of an effective spin Hall angle $\theta_{SH}^{eff}$ as $T^{\parallel} = [\hbar/(2e) \theta_{SH}^{eff}/(M_s t_{FM})]j$, where $t_{FM}$ is the thickness of the FM layer and the current density $j$ is assumed to be uniform throughout the FM/HM bilayer,~\cite{LeeAPL2013,leeAPL2014} which is a reasonable assumption for Co/Pt. Note that $\theta_{SH}^{eff}$ is a useful parameter to compare results from different experiments, but does not provide a reliable estimate of the bulk spin Hall angle of the HM layer, as it takes into account neither the finite spin diffusion length in the HM nor a possible contribution to the torque of the FM/HM interface. Here, by considering the ratio $T^{\parallel}/j= 6.9$~mT/$10^7$~Acm$^{-2}$ ($\theta_{SH}^{eff}=0.11$) obtained from harmonic Hall voltage measurements of Pt(3nm)/Co(0.6)/AlO$_x$ dots in the quasistatic, low current ($j\leq10^7$~Acm$^{-2}$) limit,~\cite{GarelloNN2013} we estimate $I_{c0} \approx 2.05$~mA. This value is about 3.5 times larger compared to the experiment. In order to match the critical current of our samples to the macrospin prediction, $\theta_{SH}^{eff}$ should be about 0.4, an unreasonably large value for Pt.~\cite{Hoffmann13IEEEtm} As $\tau_p$ is too fast for thermally assisted switching, this comparison suggests that the magnetization reverses by a more current-efficient process than coherent rotation of a single magnetic domain.\\
\indent Further support for this hypothesis comes from macrospin simulations of SOT switching in the sub-ns regime using the Landau-Lifshitz-Gilbert equation (not shown), which reveal that $I_{c} \sim \tau_{p}^{-\beta}$ with $\beta \approx 2$ rather than $\beta = 1$ as found in the experiment. This behavior differs from the macrospin dynamics of perpendicular magnetic layers induced by STT, for which our simulations confirm the linear scaling ($\beta = 1$) found in Ref.~\onlinecite{liuJMMM2014}. The difference between SOT and STT stems from the competition between $\boldsymbol{T}^{\parallel}=T^{\parallel} \boldsymbol{m} \times (\boldsymbol{m} \times \boldsymbol{y})$ and the anisotropy torque, which tend to align the magnetization respectively along $y$ and $z$ whereas in the STT case they both tend to align it towards $z$.\\
\indent The inconsistency between macrospin models and our experiment suggests that magnetization reversal occurs by domain nucleation and propagation. In such a scenario, once a reverse domain nucleates due to the antidamping and field-like SOT, switching is achieved by the propagation of a domain wall through the dot. Since the domain wall velocity is proportional to $j$, the critical switching current is expected to be proportional to $\tau_{p}^{-1}$, in agreement with our results in the short-time regime and Eq.~\eqref{eq1}. In this case, the "net charge" $q$ is inversely proportional to the domain wall velocity and can be interpreted as the angular momentum required to switch the entire dot once the reversal barrier of a portion of the sample has been overcome. The ratio between domain wall velocity and current density can be estimated as $ v/j = w /[q/S] = 137$~(m/s)$/10^{8}$~Acm$^{-2}$, where $w$ is the width of the sample and $S=w(t_{FM}+t_{HM})$ the cross section of the FM/HM bilayer. This ratio increases with increasing $B_x$, as would domain wall speed, and is in quite good agreement with the large current-induced domain wall velocities (100-400~m/s) reported on similar structures.~\cite{MironNM2011,emoriNM2013} We further note that micromagnetic simulations studies of FM/HM bilayers with large spin-orbit interaction proposed similar magnetization reversal scenarios,~\cite{finocchioAPL2013,martinezAPL2013,perezAPL2014,leePRNB2014} pointing out also the important role played by the chirality of the walls.~\cite{MironNM2011,ryuNNT2013,emoriNM2013,leePRNB2014}\\
\indent In the thermally assisted region ($\tau_{p} \gg 1$~ns), $I_c$ is predicted to be~\cite{leeAPL2014}
\begin{equation}
\label{eq2}
I_c=\frac{B_k}{4} \frac{j}{T^{\parallel}}S \left(\pi-2b_x-
\sqrt
{\begin{aligned}
	& \frac{8}{\xi} \ln{\left(\frac{-\tau_p}{\tau_0  \ln{(1-P)}} \right)}\\
	& -8-4b_x^2-4b_x (\pi-4)+\pi^2
	\end{aligned}}
\right)
\end{equation}
where $\xi = B_{k}M_{s}w^{2}t_{FM}/2k_{B}T$ is the thermal stability factor, $ b_x=B_{x}/B_k $, and $\tau_{0}$ the thermal attempt time. Although this expression is derived analytically in the framework of a macrospin model, we find that it fits reasonably well to our data (dashed line in Fig.~\ref{fig3}). The fit, performed for $\tau_p$ between 1~$\mu$s and 10~ms by taking $\tau_{0} = 1$~ns (estimated from the inflection point of the curve in Fig.~\ref{fig4}(a)), $b_x = 0.091$, and $P=0.9$, gives $\xi=110$. As for sample s3 $\xi \approx 700$ at room temperature, the smaller value of $\xi$ derived from the fit indicates that the Co layer is not reversing as a monodomain, in agreement with the conclusions drawn from the short-time regime and similar to perpendicularly magnetized nanopillars.~\cite{bedauAPL2010,bernstein2011prb} An important result from this analysis is that the intercept of the fit in the thermally assisted region (dashed line in Fig.~\ref{fig3}) and the intrinsic current determined in the short-time regime (dash-dotted line) gives the incubation time of the switching process,~\cite{bedauAPL2010,liuJMMM2014} which we find to be negligibly small ($\sim10^{-20\pm 2}$~s). Due to the weak dependence of $I_c$ on $\tau_p$ in the thermally assisted regime, this result is largely independent of the function used to fit the data.\\
\indent In conclusion, we have demonstrated non-stochastic bipolar switching of 90~nm magnetic dots induced by SOT using in-plane injection of current pulses down to 180~ps. This makes SOT-based heterostructures a promising candidate for ultra-fast recording applications such as MRAMs and cache memories. Similarly to STT, we find that the dependence of the critical switching current on the pulse length can be divided into a short-time (intrinsic) regime and a long-time (thermally assisted) regime. For $\tau_p < 1$~ns the critical switching current scales linearly with $\tau_p^{-1}$, contrary to the precessional behavior expected of a single domain magnet and consistently with a scenario where the switching speed is determined by domain wall propagation. The critical switching current is smaller than that predicted by a single domain model. In the single domain limit, the ratio between the SOT and STT critical current scales as~\cite{LeeAPL2013} $I_{c0}^{SOT}/I_{c0}^{STT} = \frac{1}{2\alpha}\frac{\eta}{\theta_{SH}^{eff}}\frac{t_{FM}+t_{HM}}{w}$, where a large spin polarization $\eta$ and low damping favor STT, whereas a large $\theta_{SH}^{eff}$ and the smaller cross section of the current injection line favor SOT. Our results indicate that ultrafast SOT switching may compare more favorably to STT when domain propagation is involved. Finally, we find that the incubation time is negligibly small, which is a very promising feature of SOT for sub-ns switching of nanomagnets.
%%%%%%%%%%%%%%%%%%%%%%%%%%%%%%%%%%%%%%%%%%%%%%%%%%%%%%%%%%%%%%%%%%%%%%%%%%%%%%%%%%%%%%%%%%%%%%%%%%%%%
\begin{acknowledgments}
This work was supported by the European Commission under the Seventh Framework Program (Grant Agreement 318144, spOt project), the Swiss National Science Foundation (Grant No. 200021-153404), the French government projects Agence Nationale de le Recherche (ANR-10-BLAN-1011-3 SPINHALL, ANR-11-BS10-0008 ESPERADO), and the European Research Council (StG 203239 NOMAD). The devices were fabricated at Nanofab-CNRS and the Plateforme de Technologie Amont (PTA) in Grenoble.
\end{acknowledgments}

\bibliographystyle{aipnum4-1}
\bibliography{UltrafastSwitchingBySOT}

\end{document}